\documentclass[12pt] {article}
\usepackage{psfrag}
\usepackage{graphicx}
\usepackage{latexsym,amsfonts}
\usepackage{amssymb}
\begin{document}
\setcounter{page}{1}
\def\theequation{\arabic{section}.\arabic{equation}}
\def\theequation{\thesection.\arabic{equation}}
\setcounter{section}{0}

\title{From Solar Proton Burning to Pionic Deuterium through the
Nambu--Jona--Lasinio model of light nuclei}

\author{A. N. Ivanov\,\thanks{E--mail: ivanov@kph.tuwien.ac.at, Tel.:
+43--1--58801--14261, Fax: +43--1--58801--14299}~\thanks{Permanent
Address: State Polytechnic University, Department of Nuclear Physics,
195251 St. Petersburg, Russian Federation}\,,
M. Cargnelli\,\thanks{E--mail: michael.cargnelli@oeaw.ac.at}\,,
M. Faber\,\thanks{E--mail: faber@kph.tuwien.ac.at, Tel.:
+43--1--58801--14261, Fax: +43--1--58801--14299}\,, H.
Fuhrmann\,\thanks{E--mail: hermann.fuhrmann@oeaw.ac.at}\,,\\
V. A. Ivanova\,\thanks{E--mail: viola@kph.tuwien.ac.at, State
Polytechnic University, Department of Nuclear Physics, 195251
St. Petersburg, Russian Federation}\,, J. Marton\,\thanks{E--mail:
johann.marton@oeaw.ac.at}\,, N. I. Troitskaya\,\thanks{State Polytechnic
University, Department of Nuclear Physics, 195251 St. Petersburg,
Russian Federation}~~, J. Zmeskal\,\thanks{E--mail:
johann.zmeskal@oeaw.ac.at}}

\date{\today}

\maketitle

\vspace{-0.5in}

\begin{center}
{\it Atominstitut der \"Osterreichischen Universit\"aten,
Arbeitsbereich Kernphysik und Nukleare Astrophysik, Technische
Universit\"at Wien, \\ Wiedner Hauptstr. 8-10, A-1040 Wien,
\"Osterreich \\ und\\ Institut f\"ur Mittelenergiephysik
\"Osterreichische Akademie der Wissenschaften,\\
Boltzmanngasse 3, A-1090, Wien, \"Osterreich}
\end{center}

\begin{center}
\begin{abstract}
Within the Nambu--Jona--Lasinio model of light nuclei (the NNJL
model), describing strong low--energy nuclear interactions, we compute
the width of the energy level of the ground state of pionic deuterium.
The theoretical value fits well the experimental data. Using the cross
sections for the reactions $\nu_e + d \to p + p + e^-$ and $\nu_e + d
\to p + n + \nu_e$, computed in the NNJL model, and the experimental
values of the events of these reactions, detected by the SNO
Collaboration, we compute the boron neutrino fluxes. The theoretical
values agree well with the experimental data and the theoretical
predictions within the Standard Solar Model by Bahcall.  We argue the
applicability of the constraints on the astrophysical factor for the
solar proton burning, imposed by helioseismology, to the width of the
energy level of the ground state of pionic deuterium. We show that the
experimental data on the width satisfy these constraints.  This
testifies an indirect measurement of the recommended value of the
astrophysical factor for the solar proton burning in terrestrial
laboratories in terms of the width of the energy level of the ground
state of pionic deuterium.
\end{abstract}


\end{center}

\newpage

\section{Introduction}
\setcounter{equation}{0}

The Nambu--Jona--Lasinio model of light nuclei
\cite{AI1}--\cite{AI4}\,\footnote{We use the abbreviation the NNJL
model that means the Nuclear Nambu--Jona--Lasinio model \cite{AI1}.}
is an attempt to describe at the quantum field theoretic level the
deuteron as a bound $np$ state \cite{FR01}. As has been shown in
\cite{AI1}--\cite{AI4} the NNJL model fits well the low--energy
parameters of the deuteron, such as the binding energy, the dipole
magnetic and electric quadrupole moments \cite{AI1}, the
$\Delta\Delta$ component \cite{AI2} and the asymptotic ratio $D/S$
\cite{AI4} of the wave function of the deuteron\,\footnote{The
asymptotic ratio $D/S$ of the D--wave component to the S--wave
component of the wave function of the deuteron in the ground state has
been computed in the NNJL model in agreement with the results obtained
by Ericson within the potential model approach \cite{TE82} and the
experimental value, which has been used by Kamionkowski and Bahcall
\cite{JB94} for the calculation of the astrophysical factor for the
solar proton burning $p + p \to d + e^+ + \nu_e$.}.

The application of the NNJL model to the description of low--energy
reactions of astrophysical interest \cite{AI3} has allowed to compute:
(i) the cross section for the neutron--proton radiative capture for
thermal neutrons $n + p \to d + \gamma$ in agreement with experimental
data with an accuracy better than $3\,\%$, (ii) the astrophysical
factor for the solar proton burning $p + p \to d + e^+ + \nu_e$,
$S_{pp}(0) = 4.08\times 10^{-25}\,{\rm MeV\,b}$, agreeing well with
the recommended value $S^{\rm SSM}_{pp}(0) = 4.00\times 10^{-25}\,{\rm
MeV\,b}$ \cite{EA98}, accepted in the Standard Solar Model (SSM) by
Bahcall \cite{JB02,JB04}, (iii) the astrophysical factor for the
reaction $p + e^- + p \to d + \nu_e$ in analytical agreement with the
result obtained by Bahcall \cite{JB64}, (iv) the cross sections for
the reactor anti--neutrino disintegration of the deuteron $\bar{\nu}_e
+ d \to n + n + e^+$ and $\bar{\nu}_e + d \to p + n + \bar{\nu}_e$,
induced by the charged and neutral weak current, respectively, in
agreement with the experimental data by the Reines Group \cite{FR99}.

In this paper we apply the NNJL model to the calculation of the width
of the ground state of pionic deuterium.  We show that the theoretical
value agrees well with the experimental data. In the NNJL model the
astrophysical factor for the solar proton burning and the width of the
ground state of pionic deuterium are defined by the same matrix
element, caused by the anomaly of the one--nucleon loop diagram. Due
to this we suggest to apply the constraints on the astrophysical
factor for the solar proton burning, imposed by helioseismology
\cite{SD98}, to the width of the energy level of the ground state of
pionic deuterium. We show that the available experimental data
\cite{PD1}--\cite{PD3} on the width of the energy level of the ground
state of pionic deuterium satisfy these constraints.

Remind that according to the SSM \cite{JB02,JB04}, the astrophysical
factor for the solar proton burning determines the temperature in the
core of the Sun. Since the solar neutrino fluxes depend strongly on
the solar core temperature \cite{AU96}, the precise knowledge of the
temperature in the core of the Sun or equivalently the astrophysical
factor for the solar proton burning is very important for the correct
definition of these fluxes \cite{JB02,JB04}.

As has been shown in \cite{SD98}, helioseismology imposes some
constraints on the astrophysical factor $S_{pp}(0)$ for the solar
proton burning relative to the recommended value $S^{\rm
SSM}_{pp}(0)$. These constraints read
\begin{eqnarray}\label{label1.1}
0.94 \le \frac{S_{pp}(0)}{S^{\rm SSM}_{pp}(0)} \le 1.18.
\end{eqnarray}
Below we argue that through the NNJL model the same constraints can be
applied to the width of the energy level of the ground state of pionic
deuterium.

In order to make this assumption more credible and to give an
additional confirmation that the NNJL model describes well strong
low--energy interactions in nuclear reactions with the deuteron, we
suggest to analyse the experimental data by the Sudbury Neutrino
Observatory (SNO) \cite{SNO3,SNO2} on the ${^8}{\rm B}$ solar neutrino
flux measured through the reactions $\nu_e + d \to p + p + e^-$ and
$\nu_e + d \to p + n + \nu_e$, caused by the charged and neutral weak
current, respectively. For this aim we use the cross sections for the
reactions $\nu_e + d \to p + p + e^-$ and $\nu_e + d \to p + n +
\nu_e$, computed within the NNJL model, and the experimental values of
the rates of the events of these reactions, detected by the SNO
Collaboration.

The paper is organized as follows. In Section 2 we compute the
${^8}{\rm B}$ solar neutrino fluxes using the cross sections for the
reactions $\nu_e + d \to p + p + e^-$ and $\nu_e + d \to p + n +
\nu_e$, computed within the NNJL model and averaged over the ${^8}{\rm
B}$ solar neutrino flux obtained by Bahcall {\it et al.}  \cite{JB96},
and the experimental values of the rates of favourable events,
detected by the SNO Collaboration. We show that the ${^8}{\rm B}$
solar neutrino flux, computed through the reaction $\nu_e + d \to p +
n + \nu_e$ and caused by the neutral weak current, fits well the
experimental data by the SNO Collaboration and the theoretical value,
predicted within the SSM by Bahcall \cite{JB04}. The obtained decrease
of the ${^8}{\rm B}$ solar neutrino flux, computed through the cross
section for the reaction $\nu_e + d \to p + p + e^-$ caused by the
charged weak current, relative to that computed through the reaction
$\nu_e + d \to p + n + \nu_e$ can be explained by neutrino
oscillations. This testifies that the NNJL model describes well strong
low--energy interactions in low--energy nuclear reactions with the
deuteron. In Section 3 we compute the width of the energy level of the
ground state of pionic deuterium within the NNJL model. We show that
the theoretical value fits well the experimental data.  In the
Conclusion we discuss the obtained results. We argue that the
constraints on the astrophysical factor for the solar proton burning,
imposed by the helioseismological data, can be applied to the width of
the energy level of the ground state of pionic deuterium. We show that
the experimental data on the width of the energy level of the ground
state of pionic deuterium satisfy these constraints.

\section{SNO data on the solar neutrino disintegration of the deuteron
within the NNJL model}
\setcounter{equation}{0}

Recently \cite{SNO3} (see also \cite{SNO2}) the SNO Collaboration has
published new experimental data on the ${^8}{\rm B}$ solar neutrino
fluxes measured through the reactions $\nu_e + d \to p + p + e^-$ and
$\nu_e + d \to p + n + \nu_e$, induced by the charged and neutral weak
current
\begin{eqnarray}\label{label2.1}
\phi^{\rm SNO}_{\rm CC}({^8}{\rm B}) &=& (1.70 \pm 0.12)\times
10^6\,{\rm cm^{-2}\,s^{-1}},\nonumber\\ \phi^{\rm SNO}_{\rm
NC}({^8}{\rm B}) &=& (4.90 \pm 0.38)\times 10^6\,{\rm
cm^{-2}\,s^{-1}},
\end{eqnarray}
where the abbreviations CC and NC mean the {\it Charged} weak {\it
Current} and the {\it Neutral} weak {\it Current}, respectively.

According to \cite{BR97}, the ${^8}{\rm B}$ solar neutrino fluxes
measured through the reactions $\nu_e + d \to p + p + e^-$ and $\nu_e
+ d \to p + n + \nu_e$ are defined by
\begin{eqnarray}\label{label2.2}
\phi({^8}{\rm B}) = 10^{-31}\,\frac{R}{\langle
\sigma(E_{\nu_e})\rangle_{{^8}{\rm B}}},
\end{eqnarray}
where $R$ is the experimentally measured rate of the favourable
events, $\langle \sigma(E_{\nu_e})\rangle_{{^8}{\rm B}}$ is the
theoretical cross section for the reaction through which the ${^8}{\rm
B}$ solar neutrino flux is measured. The cross section is averaged
over the ${^8}{\rm B}$ solar neutrino spectrum normalized to unity
\cite{JB96}.

In our case the cross sections for the reactions $\nu_e + d \to p + p
+ e^-$ and $\nu_e + d \to p + n + \nu_e$ are computed in the NNJL
model \cite{AI3} and averaged over the ${^8}{\rm B}$ solar neutrino
spectrum obtained by Bahcall {\it et al.}  \cite{JB96}.  Using the
theoretical values for the cross sections \cite{AI3}, the experimental
values of the rates of favourable events, detected by the SNO
Collaboration \cite{SNO3,SNO2}, we get
\begin{eqnarray}\label{label2.3}
\phi({^8}{\rm B})_{\rm CC} &=& (2.33 \pm 0.38)\times 10^6\,{\rm
cm^{-2}\,s^{-1}},\nonumber\\ \phi({^8}{\rm B})_{\rm NC} &=& (6.15 \pm
1.01)\times 10^6\,{\rm cm^{-2}\,s^{-1}}.
\end{eqnarray}
It is seen that the cross section for the reaction $\nu_e + d \to p +
n + \nu_e$, computed within the NNJL model, fits well the experimental
data by the SNO Collaboration on the ${^8}{\rm B}$ solar neutrino flux
$\phi^{\rm SNO}_{\rm NC}({^8}{\rm B}) = (4.90 \pm 0.38)\times
10^6\,{\rm cm^{-2}\,s^{-1}}$.

We would like to emphasize that the ${^8}{\rm B}$ solar neutrino flux
$\phi({^8}{\rm B})_{\rm NC} = (6.15 \pm 1.01)\times 10^6\,{\rm
cm^{-2}\,s^{-1}}$ agrees also well with the theoretical ${^8}{\rm B}$
solar neutrino flux, predicted within the SSM by Bahcall \cite{JB04}:
$\phi^{\rm SSM}({^8}{\rm B}) = (5.82 \pm 1.34)\times 10^6\,{\rm
cm^{-2}\,s^{-1}}$.

The cross section for the reaction $\nu_e + d \to p + p + e^-$,
induced by the charged weak current and computed within the NNJL
model, leads to the theoretical prediction for the observed ${^8}{\rm
B}$ solar neutrino flux, measured through the reaction $\nu_e + d \to
p + p + e^-$, agreeing with the experimental value $\phi^{\rm
SNO}_{\rm CC}({^8}{\rm B}) = (1.70 \pm 0.12)\times 10^6\,{\rm
cm^{-2}\,s^{-1}}$ within two standard deviations but by a factor of 3
smaller compared with the ${^8}{\rm B}$ solar neutrino flux, measured
through the reaction $\nu_e + d \to p + n + \nu_e$, induced by the
neutral weak current.

According to the generally accepted point of view, such a distinction
can be explained by neutrino oscillations \cite{SNO69,SNO99} (see also
\cite{JB04}). Remind that the cross section for the reaction $\nu_X +
d \to p + n + \nu_X$ is practically insensitive to the neutrino
flavour $X = e ,\mu$ or $\tau$ \cite{JB04,SNO99}.

The obtained results testify that the NNJL model describes well strong
low--energy interactions in low--energy nuclear reactions with the
deuteron.

\section{Width of the energy level of the ground state of pionic 
deuterium}
\setcounter{equation}{0}

According to Deser, Goldberger, Baumann and Thirring \cite{DT54,TE88}
(see also \cite{IV1}--\cite{IV4}), the width of the energy level of
the ground state of pionic deuterium is defined by the DGBT formula
\begin{eqnarray}\label{label3.1}
\Gamma_{1s} = 4\alpha^3 m^2_{\pi}\,{\cal I}m\, f^{\pi^-d}_0(0),
\end{eqnarray}
where $\alpha = e^2 = 1/137.036$ is the fine structure constant in
Gaussian units and $m_{\pi} = 140\,{\rm MeV}$ is the pion mass,
$f^{\pi^-d}_0(0)$ is the S--wave amplitude of $\pi^-d$ scattering near
threshold.

For the analyses of the imaginary part of the amplitude
$f^{\pi^-d}_0(0)$ it is sufficient to take into account the
contribution of two processes, $\pi^- d \to n n$ and $\pi^- d \to n n
\gamma$, only. This defines the width (\ref{label3.1}) as follows
\begin{eqnarray}\label{label3.2}
\Gamma_{1s}= 4\alpha^3\,m^2_{\pi}\,({\cal
I}m\,f^{\pi^-d}_0(0)_{nn\gamma} + {\cal I}m\,f^{\pi^-d}_0(0)_{nn}) =
\Gamma^{(nn\gamma)}_{1s} + \Gamma^{(nn)}_{1s},
\end{eqnarray}
where ${\cal I}m\,f^{\pi^-d}_0(0)_{nn\gamma}$ and ${\cal
I}m\,f^{\pi^-d}_0(0)_{nn}$ are the imaginary parts of the S--wave
amplitudes of $\pi^-d$ scattering near threshold saturated by the
intermediate $nn\gamma$ and $nn$ states, and
$\Gamma^{(nn\gamma)}_{1s}$ and $\Gamma^{(nn)}_{1s}$ are the partial
widths of the decays $A_{\pi d} \to nn\gamma$ and $A_{\pi d} \to nn$,
respectively.

Following \cite{IV1}--\cite{IV4} (see also \cite{TE04}) the S--wave
amplitudes $f^{\pi^-d}_0(0)_{nn\gamma}$ and $f^{\pi^-d}_0(0)_{nn}$ can
be defined by
\begin{eqnarray}\label{label3.3}
\hspace{-0.3in}&&f^{\pi^-d}_0(0)_{nn\gamma} =
\frac{1}{8\pi}\,\frac{1}{m_d +
m_{\pi}}\,\frac{\alpha}{F^2_{\pi}}\,\frac{1}{2}\int \frac{d^3p}{(2\pi)^2
2|\vec{p}\,|}\int \frac{d^3k_1}{(2\pi)^3
2E_n(k_1)}\frac{d^3k_2}{(2\pi)^3 2E_n(k_2)}\nonumber\\
\hspace{-0.3in}&&\times\,(2\pi)^3\,\delta^{(3)}(\vec{p} - \vec{k}_1
- \vec{k}_2)\,\frac{1}{E_n(k_1) + E_n(k_2) + |\vec{p}\,| - m_{\pi} -
m_d - \,i\,0}\nonumber\\
\hspace{-0.3in}&&\times\,\frac{1}{3}\sum_{\alpha_2 = \pm 1/2}
\sum_{\alpha_1 = \pm 1/2} \sum_{\lambda_d = 0,\pm 1}\sum_{\lambda =
\pm 1}|e^{*\mu}(p,\lambda) \langle n(\vec{k}_1,\alpha_1)
n(\vec{k}_2,\alpha_2)|
J^{1-i2}_{5\mu}(0)|d(\vec{0},\lambda_d)\rangle|^2
\end{eqnarray}
and 
\begin{eqnarray}\label{label3.4}
\hspace{-0.3in}&&f^{\pi^-d}_0(0)_{nn} = \frac{1}{128\pi}\,\frac{1}{m_d
+ m_{\pi}}\,\int \frac{d^3k}{(2\pi)^3 E^2_n(k)}\,\frac{1}{E_n(k) - m_N
- m_{\pi}/2 - \,i\,0}\nonumber\\
\hspace{-0.3in}&&\times\,\frac{1}{3}\sum_{\alpha_2 = \pm 1/2}
\sum_{\alpha_1 = \pm 1/2} \sum_{\lambda_d = 0,\pm
1}|M(\pi^-(\vec{0}\,)d(\vec{0},\lambda_d) \to n(\vec{k},\alpha_1) n(-
\vec{k},\alpha_2))|^2.
\end{eqnarray}
In (\ref{label3.2}) the matrix element of the transition $\pi^-d \to
 nn \gamma$ is given in the soft--pion limit
 \cite{IV3,TE04}--\cite{TE77}. According to the Pauli principle
 \cite{MF02} the $nn$ pair in the reaction $\pi^-d \to nn$, where
 $\pi^-d$ pair is in the S--wave state, can be only in the ${^3}{\rm
 P}_1$ state.

Computing the matrix element of the axial--vector current and the
amplitude of the reaction $\pi^-d \to nn$ in the NNJL model, for the
partial widths of the decays $A_{\pi d} \to nn \gamma$ and $A_{\pi d}
\to nn$ we obtain
\begin{eqnarray}\label{label3.5}
\Gamma^{(nn\gamma)}_{1s} = m^2_{\pi} m^2_N\,
g^2_V\,C^2_{NN}\,\frac{3\alpha^4}{64
\pi^7}\,\frac{g^2_A}{F^2_{\pi}}\,\int^{\infty}_0 \frac{dk\,k^2
F^2_d(k^{\,2})}{\displaystyle \Big(1 -
\frac{1}{2}\,r_{nn}a_{nn}k^2\Big)^2 + a^2_{nn}k^2}.
\end{eqnarray}
and
\begin{eqnarray}\label{label3.6}
\Gamma^{(nn)}_{1s} =
\alpha^3\,C^2_{NN}\,\frac{m^4_{\pi}}{F^2_{\pi}}\,\frac{3g^2_A
g^2_V}{256\pi^7}\,k^3_0\,F^2_d(k^2_0)\,|f^{(nn;{^3}{\rm
P}_1)}_{\pi^-d}(k_0)|^2,
\end{eqnarray}
where $g_A = 1.267$ is the axial--vector coupling constant, $F_{\pi} =
92.4\,{\rm MeV}$ is the leptonic constant of charged pions, $g_V =
11.3$ and $C_{NN} = 3.27\times 10^{-3}\,{\rm MeV}^{-2}$ are the
coupling constants of the NNJL model, $F_d(k^2) = 1/(1 + r^2_d k^2)$
is the form factor of the deuteron, proportional to the wave function
of the ground state of the deuteron in the momentum representation,
$r_d = 4.32\,{\rm fm} = 3.07\,{\rm m^{-1}_{\pi}}$ is the deuteron
radius \cite{MN79}, $a_{nn}$ and $r_{nn}$ are the S--wave scattering
length of the $nn$ scattering in the ${^1}{\rm S}_0$ state. For
numerical calculation we use $a_{nn} = -23.75\,{\rm fm} = -
16.85\,{\rm m^{-1}_{\pi}}$ and $r_{nn} = 2.75\,{\rm fm} = 1.95\,{\rm
m^{-1}_{\pi}} $ \cite{AI3}. Then, the relative momentum $k_0$ of the
$nn$ pair at threshold of the reaction $\pi^-d \to nn$ in the center
of mass frame is equal to $k_0 = \sqrt{m_{\pi} m_N} = 362\,{\rm MeV}$.
The amplitude $f^{(nn;{^3}{\rm P}_1)}_{\pi^-d}(k_0)$ describes the
final--state interaction of the $nn$ pair in the ${^3}{\rm P}_1$ state
near threshold of the reaction $\pi^-d \to nn$. Following \cite{IV4}
we compute $|f^{(nn;{^3}{\rm P}_1)}_{\pi^-d}(k_0)| = 0.7$.

In (\ref{label3.5}) the integral over $k$ amounts to $0.016/r^3_d$.
The theoretical values of the partial widths of the decays $A_{\pi d}
\to nn \gamma$ and $A_{\pi d} \to nn$ read
\begin{eqnarray}\label{label3.7}
\Gamma^{(nn\gamma)}_{1s} &=&(0.30 \pm 0.04)\,{\rm eV},\nonumber\\
\Gamma^{(nn)}_{1s} &=& (0.85 \pm 0.11)\,{\rm eV}.
\end{eqnarray}
According to (\ref{label3.2}), for the total width of the energy level
of the ground state of pionic deuterium we get
\begin{eqnarray}\label{label3.8}
\Gamma_{1s} = (1.15 \pm 0.12)\,{\rm eV}.
\end{eqnarray}
Our theoretical value of the width $\Gamma_{1s} = (1.15 \pm
0.12)\,{\rm eV}$ agrees well with the experimental data 
\begin{eqnarray}\label{label3.9}
\Gamma^{\exp}_{1s} = \left\{\begin{array}{r@{\quad,\quad}l} (1.02\pm
0.21)\,{\rm eV} & \cite{PD1,PD2}\\ (1.19 \pm 0.11)\,{\rm eV} &
\cite{PD3}.
\end{array}\right.
\end{eqnarray}
The partial widths $\Gamma^{(nn\gamma)}_{1s}$ and $\Gamma^{(nn)}_{1s}$
can be also related by the parameter $D$:
\begin{eqnarray}\label{label3.10}
D = \frac{\sigma(\pi^-d \to n n)}{\sigma(\pi^-d \to n n \gamma)} =
 \frac{\Gamma^{(nn)}_{1s}}{\Gamma^{(nn\gamma)}_{1s}} = 2.83 \pm 0.04,
\end{eqnarray}
measured experimentally at threshold of the reactions $\pi^-d \to n n$
and $\pi^-d \to n n \gamma$ \cite{D1}. Using the theoretical values of
the partial widths (\ref{label3.7}) we compute the parameter $D$: $D =
2.83 \pm 0.50$. It agrees well with the experimental data
(\ref{label3.10}).

\section{Conclusion}
\setcounter{equation}{0}

We have applied the NNJL model to the calculation of the width of the
energy level of the ground state of pionic deuterium. Without
introduction of new input parameters we have computed the value of the
width of the energy level of the ground state of pionic deuterium
$\Gamma_{1s} = (1.15 \pm 0.12)\,{\rm eV}$ in complete agreement with
the experimental data (\ref{label3.9}).  

Remind that the NNJL model has been invented for the quantum field
theoretic description of the deuteron as a bound $np$ state and
low--energy nuclear reactions with the deuteron of the astrophysical
interest such as the solar proton burning and so on. However, as has
turned out the NNJL model can be also applied to the calculation of
the width of the energy level of the ground state of pionic deuterium,
since the amplitudes of the solar proton burning $p + p \to d + e^+ +
\nu_e$, the $pep$ reaction $p + e^- + p \to d + \nu_e$, the neutrino
disintegration of the deuteron $\nu_e + d \to p + p + e^-$ and $\nu_e
+ d \to p + n + \nu_e$ and the reactions $\pi^- + d \to n + n +
\gamma$ and $\pi^- + d \to n + n $ near threshold of the $\pi^-d$ pair
are defined by the anomaly of the same one--nucleon loop diagram
\cite{AI3}.

Since in the SSM the astrophysical factor of the solar proton burning
is related to the temperature of the solar core, the
helioseismological data become sensitive to the value of the
astrophysical factor for the solar proton burning.  The constraints on
the value of the astrophysical factor for the solar proton burning,
coming from the helioseismological data on the values of sound speed
and density inside the Sun, have been found by Degl'Innocenti,
Fiorentini and Ricci \cite{SD98}.

Since the NNJL model fits well the recommended value of the
astrophysical factor for the solar proton burning and the experimental
data on the width of the energy level of the ground state of pionic
deuterium, one can imagine that the constraints on the astrophysical
factor for the solar proton burning, imposed by helioseismology
(\ref{label1.1}), can be also valid for the width of the energy level
of the ground state of pionic deuterium. This yields
\begin{eqnarray}\label{label4.1}
(1.08 \pm 0.11)\,{\rm eV} \le \Gamma_{1s} \le (1.36 \pm 0.14)\,{\rm
eV}.
\end{eqnarray}
It is seen that the experimental data (\ref{label3.9}) satisfy well
the constraints (\ref{label4.1}).

Moreover, since the astrophysical factor for the solar proton burning,
$S_{pp}(0) = 4.08 \times 10^{-25}\,{\rm MeV\,b}$, computed within the
NNJL model, fits the recommended value $S^{\rm SSM}_{pp}(0) = 4.00
\times 10^{-25}\,{\rm MeV\,b}$ with an accuracy about $2\,\%$, our
prediction for the width of the energy level of the ground state of
pionic deuterium, agreeing with the experimental data with an accuracy
about $3\,\%$, can be valued as an indirect measurement of the
recommended value of the astrophysical factor $S^{\rm SSM}_{pp}(0) =
4.00 \times 10^{-25}\,{\rm MeV\,b}$ in terrestrial laboratories in
terms of the width of the energy level of the ground state of pionic
deuterium.

For the confirmation of the applicability the NNJL model to the
description of strong low--energy interactions with the deuteron and
the results obtained above we have analysed the experimental data on
the ${^8}{\rm B}$ solar neutrino flux measured by the SNO
Collaborations. Using the cross sections for the reactions $\nu_e + d
\to p + p + e^-$ and $\nu_e + d \to p + n + \nu_e$, computed within
the NNJL model and averaged over the ${^8}{\rm B}$ solar neutrino
spectrum by Bahcall {\it et al.} \cite{JB96}, and the experimental
values of the rates of the events of the reactions $\nu_e + d \to p +
p + e^-$ and $\nu_e + d \to p + n + \nu_e$, detected by the SNO
Collaboration, we have computed the ${^8}{\rm B}$ solar neutrino
fluxes.

The computed value $\phi({^8}{\rm B})_{\rm NC} = (6.15 \pm 1.01)\times
10^6\,{\rm cm^{-2}\,s^{-1}}$ of the ${^8}{\rm B}$ solar neutrino flux,
measured through the reaction $\nu_e + d \to p + n + \nu_e$, agrees
well with the experimental data and the theoretical value of the
${^8}{\rm B}$ solar neutrino flux $\phi^{\rm SSM}({^8}{\rm B}) = (5.82
\pm 1.34)\times 10^6\,{\rm cm^{-2}\,s^{-1}}$, predicted within the SSM
by Bahcall \cite{JB04}.

In turn, the computed value $\phi({^8}{\rm B})_{\rm CC} = (2.33 \pm
0.38)\times 10^6\,{\rm cm^{-2}\,s^{-1}}$ of the ${^8}{\rm B}$ solar
neutrino flux, measured through the reaction $\nu_e + d \to p + p +
e^-$, agrees with the experimental data within two standard deviations
but differs by a factor of 3 from the ${^8}{\rm B}$ neutrino flux
$\phi({^8}{\rm B})_{\rm NC} = (6.15 \pm 1.01)\times 10^6\,{\rm
cm^{-2}\,s^{-1}}$. However, nowadays there is a consensus
\cite{JB04,SNO99} that such a distinction can be explained by solar
neutrino oscillations.

Such an agreement of the computed ${^8}{\rm B}$ solar neutrino fluxes
with the experimental data by the SNO Collaboration and the
theoretical predictions of the SSM by Bahcall \cite{JB04} testify that
the NNJL model describes well strong low--energy interactions in
low--energy nuclear reactions with the deuteron. 

This makes also credible our assumption concerning the applicability
of the constraints on the solar proton burning, coming from
helioseismology, to the width of the ground state of pionic deuterium
and vice versa.

\newpage

\section*{Acknowledgement}

We are grateful to Torleif Ericson, Heinz Oberhummer and Walter Grimus
for fruitful discussions and John Bahcall for encouraging remarks.

Natalia Troitskaya thanks the Atom Institute of the Austrian
Universities at Vienna University of Technology for kind hospitality
extended to her during the period of the work under this paper.


\begin{thebibliography}{9}
\bibitem{AI1}
A. N. Ivanov, N. I. Troitskaya, H. Oberhummer, and M. Faber,
Eur. Phys. J.  A {\bf 7}, 519 (2000), nucl-th/0006049.
\bibitem{AI2}
A. N. Ivanov, N. I. Troitskaya, H. Oberhummer, and M. Faber,
Eur. Phys. J.  A {\bf 8}, 129  (2000), nucl--th/0006050.
\bibitem{AI3}
A. N. Ivanov, N. I. Troitskaya, H. Oberhummer, and M. Faber,
Eur. Phys. J.  A {\bf 8}, 223  (2000), nucl--th/0006051.
\bibitem{AI4}
A. N. Ivanov, V. A. Ivanova, H. Oberhummer, N. I. Troitskaya and
M. Faber, 
Eur. Phys. J. A {\bf 12}, 87 (2001, nucl--th/0108067.
\bibitem{FR01}
X. Kong and F. Ravndal,
Phys. Rev. C {\bf 64}, 044002 (2001), nucl-th/0004038.
\bibitem{TE82}
T. E. O. Ericson and M. Rosa--Clot,
Phys. Lett. B {\bf 110}, 193 (1982);
T. E. O. Ericson,
Nuovo Cimento, {\bf 76}, 277 (1983);
T. E. O. Ericson and M. Rosa--Clot,
Nucl. Phys. A {\bf 405}, 497 (1983).
\bibitem{JB94}
M. Kamionkowski and J. N. Bahcall,
Astrophys. J. {\bf 420}, 884 (1994).
\bibitem{EA98}
E. G. Adelberger {\it et al.},
Rev. Mod. Phys. {\bf 70}, 1265 (1998).
\bibitem{JB02}
J. N. Bahcall,
Nuclear Physics B (Proc. Suppl.), {\bf 118}, 77 (2002), 
astro-ph/0209080;
J. N. Bahcall, M. H. Pinsonneault, and S. Basu,
Astrophys. J. {\bf 555}, 990 (2001);
J. N. Bahcall and M. H. Pinsonneault,
Phys. Rev. Lett. {\bf 92}, 121301 (2004);
http://www.sns.ias.edu/~jnb/
\bibitem{JB04}
J. N. Bahcall and C. Pe$\tilde{\rm n}$a--Garay,
New Journal of Physics, {\bf 6}, 63 (2004).
\bibitem{JB64}
J. N. Bahcall,
Astrophys.J. {\bf 139}, 318 (1964);
J. N. Bahcall and R. M. May,
Astrophys.J. {\bf 155}, 501 (1969).
\bibitem{FR99}
S. P. Riley, Z. D. Greenwood, W. R. Kroop, L. R. Price,
F. Reines, H. W. Sobel, Y. Declais, A. Etenko, and
M. Skorokhvatov,
Phys. Rev. C {\bf 59}, 1780 (1999).
\bibitem{SD98}
S. Degl'Innocenti, G. Fiorentini and B. Ricci,
Phys. Lett. B {\bf 416}, 365 (1998).
\bibitem{PD1}
D. Chatellard {\it et al.} (the PSI Collaboration),
Phys. Rev. Lett. {\bf 74}, 4157 (1995).
\bibitem{PD2}
D. Chatellard {\it et al.} (the PSI Collaboration),
Nucl. Phys. A {\bf 625}, 855 (1997).
\bibitem{PD3}
P. Hauser {\it et al.} (the PSI Collaboration),
Phys. Rev. C {\bf 58}, R1869 (1998). 
\bibitem{AU96}
J. N. Bahcall and A. Ulmer, Phys. Rev. D {\bf 8}, 4202
(1996).
\bibitem{SNO3}
S. N. Ahmed {\it et al.} (the SNO Collaboration),
Phys. Rev. Lett. {\bf 92}, 181301 (2004).
\bibitem{SNO2}
Q. R. Ahmed {\it et al.} (the SNO Collaboration),
Phys. Rev. Lett. {\bf 89}, 011301 (2002).
\bibitem{JB96}
J. N. Bahcall, E. Lisi, D. E. Alburger, L. De Braeckeleer, 
S. J. Freedmann, and J. Napolitano,
Phys. Rev. C {\bf 54}, 411 (1996).
\bibitem{BR97}
J. N. Bahcall,
in {\it NEUTRINO ASTROPHYSICS}, Cambridge University Press,
Cambridge, 1989;
V. Castellani, S. Degl'Innocenti, G. Fiorentini, M. Lissia, 
and B. Ricci,
Phys. Rep. {\bf 281}, 309 (1997).
\bibitem{SNO69}
V. N. Gribov and B. M. Pontecorvo,
Phys. Lett. B {\bf 28}, 493 (1969);
J. N. Bahcall and S. C. Frautschi, 
Phys. Lett. B {\bf  29}, 623 (1969).
\bibitem{SNO99}
S. M. Bilenky, C. Giunti, and W. Grimus,
Prog. Part. Nucl. Phys. {\bf 43},1 (1999), hep-ph/9812360]; 
J. N. Bahcall, P. I. Krastev, and A. Yu. Smirnov,
JHEP {\bf 0105}, 015 (2001);
J. N. Bahcall, M. C. Gonzalez-Garcia, and C. 
Pe$\tilde{\rm n}$a--Garay,
JHEP {\bf 0108}, 014 (2001);
K. Fujii, C. Habe, and T. Yabuki,  
Phys. Rev. D {\bf 64}, 
013011 (2001); 
J. N. Bahcall and C. Pe$\tilde{\rm n}$a--Garay,
JHEP {\bf 0311}, 004 (2003);
M. C. Gonzalez-Garcia and Y. Nir,
Rev. Mod. Phys. {\bf 75}, 345 (2003), hep-ph/0202058];
S. Pakvasa and J.W.F. Valle,
Proc. Indian National Acad. Sci. A {\bf 70}, 189 (2003), 
hep-ph/0301061;
W. Grimus,
Lectures given at the
\textit{41. Internationale Universit\"atswochen f\"ur 
Theoretische Physik, Flavour Physics},
Schladming, Styria, Austria, 22--28 February 2003; hep-ph/0307149; 
V. Barger, D. Marfatia, and K. Whisnant,
Int. J. Mod. Phys. E {\bf 12}, 569 (2003), hep-ph/0308123; 
A. Yu. Smirnov,
Invited talk given at the \textit{21st International Symposium on 
Lepton and Photon Interactions at High Energies (LP 03)},
Batavia, Illinois, U.S.A., 11--16 August 2003,
Int. J. Mod. Phys. A {\bf 19}, (2004) 1180; hep-ph/0311259;
K. Fujii and T. Shimomura, hep--ph/0212076, hep--ph/0402274, 
hep--ph/0406079.
\bibitem{DT54} 
S. Deser, M. L. Goldberger, K. Baumann, and
W. Thirring, Phys. Rev. {\bf 96}, 774 (1954);
T. L. Trueman,
Nucl. Phys. {\bf 26}, 57 (1961).
\bibitem{TE88} 
T. E. O. Ericson and W. Weise, 
in {\it PIONS AND NUCLEI}, Clarendon Press, Oxford, 1988.
\bibitem{IV1} 
A. N. Ivanov, M. Faber, A. Hirtl, J. Marton, and N. I. Troitskaya,
Eur. Phys. J. A {\bf 18}, 653 (2003), nucl--th/0306047.
\bibitem{IV2} 
A. N. Ivanov, M. Faber, A. Hirtl, J. Marton, and N. I. Troitskaya,
Eur. Phys. J. A {\bf 19}, 413 (2004) nucl--th/0310027.
\bibitem{IV3}
A. N. Ivanov, M. Cargnelli, M. Faber, J. Marton, 
N. I. Troitskaya, and J. Zmeskal,
EPJA (2004), DOI 10.1140/epja/i2003--10178--y, nucl--th/0310081. 
\bibitem{IV4} 
A. N. Ivanov, M. Faber, M. Cargnelli, H. Fuhrmann, V. A. Ivanova, 
J. Marton, N. I. Troitskaya, and J. Zmeskal,
{\it On kaonic deuterium. Quantum field theoretic and relativistic
covariant approach}, nucl--th/0406053 (to appear in EPJA).
\bibitem{TE04}
T. E. O. Ericson and A. N. Ivanov,
{\it Energy shift in pionic hydrogen from radiative processes},
(in progress).
\bibitem{SL66}
S. L. Adler and Y. Dothan,
Phys. Rev. {\bf 151}, 1267 (1966).
\bibitem{SA68}
S. L. Adler and R. Dashen,
in {\it CURRENT ALGEBRAS}, Benjamin, New York 1968.
\bibitem{ME72}
M. Ericson and M. Rho,
Phys. Rep. {\bf 5}, 57 (1972).
\bibitem{HP73}
V. De Alfaro, S. Fubini, G. Furlan, and C. Rossetti,
in {\it CURRENTS IN HADRON PHYSICS},
North--Holland, Amsterdam, 1973.
\bibitem{TE77}
J. Bernab${\acute{\rm e}}$u, T. E. O. Ericson, and C. Jarglskog,
Phys. Lett. B {\bf 69}, 161 (1977);
T. E. O. Ericson and J. Bernab${\acute{\rm e}}$u,
Phys. Lett. B {\bf 70}, 170 (1977).
\bibitem{MF02}
M. Faber, in Lecture notes on
{\it Atomic, Nuclear and Particle Physics}. II,
Vienna University of Technology, 2002.
\bibitem{MN79}
M. M. Nagels {\it et al.},
Nucl. Phys. B {\bf 147}, 179 (1979).
\bibitem{D1}
V. L. Highland, M. Salomon, M. D. Hasinoff, E. Mazzucato, 
D. F. Measday, J.--M. Poutissou, and T. Suzuki,
Nucl. Phys. A {\bf 365}, 333 (1981).
\end{thebibliography}
\end{document}